\documentclass[%
 reprint,
superscriptaddress,
amsmath,amssymb,
aps,
prl,
]{revtex4-1}

\usepackage{graphicx}% Include figure files
\usepackage{dcolumn}% Align table columns on decimal point
\usepackage{bm}% bold math
\begin{document}

\preprint{APS/123-QED}

\title{A New Electric Field in Asymmetric Magnetic Reconnection}

%\title{The Larmor Electric Field: An Electric Field Structure from Kinetic Physics of Collisionless Asymmetric Magnetic Reconnection}% Force line breaks with \\

\author{K.~Malakit}
\affiliation{
Department of Physics, Mahidol University, Bangkok 10400, Thailand
}
\affiliation{
Department of Physics and Astronomy, University of Delaware, Newark, Delaware 19716, USA
}
%\altaffiliation[Also at ]{
%Department of Physics and Astronomy, University of Delaware, Newark, Delaware 19716, USA
%}
%\email{kmalakit@gmail.com}

\author{M.~A.~Shay}
\affiliation{
Department of Physics and Astronomy, University of Delaware, Newark, Delaware 19716, USA
}

\author{P.~A.~Cassak}
\affiliation{
 Department of Physics, West Virginia University, Morgantown, West Virginia 26506, USA
}

%\author{M.~M.~Kuznetsova}
%\affiliation{
%NASA Goddard Space Flight Center, Greenbelt, Maryland 20771, USA
%}

\author{D.~Ruffolo}
\affiliation{
Department of Physics, Mahidol University, Bangkok 10400, Thailand
}
\affiliation{
Thailand Center of Excellence in Physics, CHE, Ministry of Education, Bangkok 10400, Thailand
}


%\affiliation{
%\affiliation{%


\date{\today}% It is always \today, today,

\begin{abstract}

%  It has been known that during collisionless reconnection, there exists
%  the Hall electric field structure in the plane of reconnection.
%  In this work, we theoretically argue and show numerical evidence from
%  fully-kinetic particle-in-cell simulations that there can exist another
%  in-plane electric field structure pointing away from the X-line when the
%  inflow conditions are so asymmetric that the ion Larmor radius exceeds
%  the distance from the edge of the dissipation region to the stagnation point
%  This electric field, dubbed the ``Larmor electric field,'' is expected
%  to be present at the dayside magnetopause reconnection site, appearing
%  on the magnetospheric side and pointing Earthward.  This electric field
%  could be used as a signature of the dissipation region on the
%  dayside, which could be important for locating reconnection sites by
%  the upcoming Magnetospheric Multiscale (MMS) mission.
%

  We present a theory and numerical evidence for the existence of
  a previously unexplored in-plane electric field in collisionless
  asymmetric magnetic reconnection. This electric field,
  dubbed the ``Larmor electric field,'' is associated with finite
  Larmor radius effects and is distinct from the known Hall electric field.
  Potentially, it could be an important indicator
  for the upcoming Magnetospheric Multiscale (MMS) mission to
  locate reconnection sites as we expect it to appear on the magnetospheric
  side, pointing Earthward, at the dayside magnetopause reconnection site.

%  Theoretical and numerical evidence for the existence of an in-plane
%  electric field in collisionless asymmetric magnetic reconnection is
%  presented.  The electric field is associated with finite Larmor
%  radius effects and is distinct from the electric field associated
%  with the Hall effect.  In asymmetric reconnection, the X-line and
%  stagnation point are not necessarily colocated.  If the Larmor
%  radius on one side of the dissipation region exceeds the distance
%  from the edge of the dissipation region to the stagnation point, the
%  electric field is present and points away from the X-line.  A basic
%  theoretical argument for its existence is offered, and fully kinetic
%  particle-in-cell simulations with various asymmetric inflow
%  conditions are used to verify the predictions.  This electric field,
%  dubbed the ``Larmor electric field,'' is expected to be present at
%  the dayside magnetopause reconnection site, appearing on the
%  magnetospheric side and pointing Earthward.  This electric field
%  could be used as a signature of the dissipation region on the
%  dayside, which could be important for locating reconnection sites by
%  the upcoming Magnetospheric Multiscale (MMS) mission.

\begin{description}
\item[PACS numbers] 52.35.Vd, 94.30.cp
%\pacs{52.35.Vd}% PACS, , 94.30.cp the Physics and Astronomy
\end{description}
\end{abstract}

\maketitle

%\tableofcontents

\section{\label{sec:intro}Introduction}

%{\bf Introduction}
Magnetic reconnection efficiently converts magnetic energy into heat
and flow energy of particles in plasmas ({\it e.g.,} \cite{Birn07}).
It occurs when small-scale dissipation permits an electric field that
breaks the frozen-in condition and allows magnetic field lines to
change topology.  It naturally arises in many locations in the Earth's
magnetosphere.  The dissipation region is often difficult to measure
in naturally occurring settings because it is small compared to global
scales.  The primary objective of the upcoming Magnetospheric
MultiScale (MMS) mission is to study the properties of the dissipation
region in reconnection \cite{Burch09,Moore12}.  Therefore, it is of
critical importance to determine the observational signatures of
magnetic reconnection near the dissipation region.

One signature of the dissipation region in collisionless reconnection
is due to the Hall effect.  Since ions and electrons have different
Larmor radii due to their different masses, they undergo different
bulk motion within small distances from the reconnection site.  This
sets up in-plane Hall currents, which are associated with out-of-plane
magnetic fields and in-plane electric fields
\cite{Sonnerup79,Terasawa83,Mandt94,Shay98a}.  During symmetric reconnection
as is most often studied theoretically and numerically, the Hall
magnetic field is quadrupolar and the Hall electric field is bipolar.
This signature of collisionless reconnection has been observed using
satellite observations at the dayside magnetopause
\cite{Nagai01,Oieroset01,Mozer02,Scudder02,Runov03}.  Note that
observations of the quadrupolar Hall magnetic and bipolar Hall
electric field are rare at the dayside magnetopause \cite{Mozer07}
because reconnection normally has asymmetric inflow, which changes the
structure of the Hall fields \cite{Karimabadi99,Swisdak03}.  In some
cases, the asymmetry causes the Hall magnetic and electric fields to
become so skewed that they become bipolar and unipolar, respectively
\cite{Mozer08b,Pritchett08,Tanaka08,Malakit10}.  These skewed Hall
structures have also been observed \cite{Mozer08a,Mozer08b,Tanaka08}.

In this paper, we argue for the existence of an in-plane electric field
that was not previously discussed, which appears in asymmetric reconnection and
can be used as an observational signature of the dissipation region.
This electric field is caused by finite Larmor radius effects, so we
dub it the ``Larmor electric field.''  The electric field structure is
located in the inflow region of the dissipation region, just upstream
of the Hall electric field structure, with its direction pointing away
from the X-line.  We present a physical argument for its existence and
show it is consistent with the results of fully-kinetic
particle-in-cell (PIC) simulations.

Under normal dayside reconnection conditions, we show that the Larmor
electric field should be present on the magnetospheric side of the
X-line pointing Earthward with a magnitude large enough to be
measurable with spacecraft.  Therefore, its existence can be useful
for locating dissipation regions in the MMS mission.

\section{\label{sec:theory}Theory}

%{\bf Theory}
To understand where the Larmor electric field comes from, consider the
structure of the dissipation region.  In symmetric anti-parallel
magnetic reconnection, the X-line and stagnation point are located at
the due center of the dissipation region.  When there are asymmetries
in the magnetic field $B$ and/or plasma density $n$, this is no longer
the case \cite{Cassak07d}.
The X-line location is determined by energy conservation, while the
stagnation point is determined by mass conservation.  In particular,
the stagnation point location is given by \cite{Cassak07d}
\begin{equation}
  \frac{\delta_{S2}}{\delta_{S1}} \sim \frac{n_2B_1}{n_1B_2},
  \label{eqn:ds2ds1}
\end{equation}
where $\delta_{S}$ is the distance from the stagnation point to one
edge of the dissipation region and the subscripts ``1'' and ``2''
denote the inflow side of interest.  The stagnation point is offset
towards the side with smaller $n / B$.

\begin{figure}
  \centering{}\includegraphics[width=3.1in]{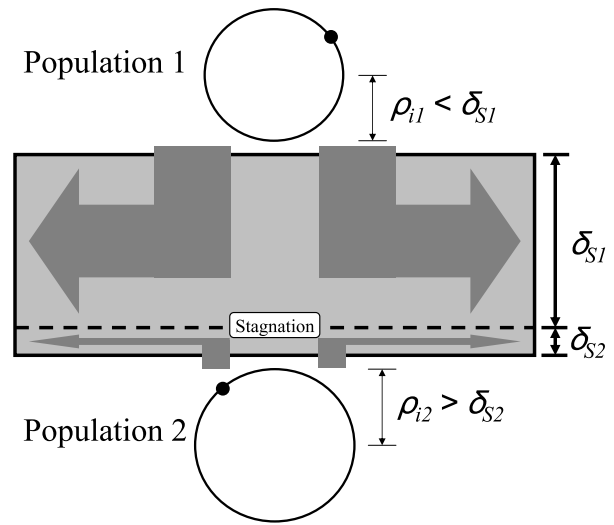}
  \caption{\label{fig:schematic} Schematic of the flow structure of
    the dissipation region during asymmetric reconnection
    for characteristic dayside-magnetosphere parameters in a kinetic plasma.}
\end{figure}

The basic picture of the dissipation region structure is sketched in
Fig.~\ref{fig:schematic}.  The sketch is roughly for typical magnetopause inflow
conditions (density variation by a factor of 10, magnetic field
variation by a factor of 2), with the magnetosheath and magnetosphere
as Populations 1 and 2, respectively.  The low magnetospheric density
implies that the stagnation point is shifted very close to the Population~2
side of the dissipation region. The inflowing Population~2 plasma is
therefore constrained in a basic fluid sense to turn the corner
sharply and flow outwards.  Once magnetic field lines reconnect,
however, the two populations begin to mix along the magnetic field
lines.

A question arises in the context of a kinetic plasma.  The Larmor
radii of the two populations are shown as circles in the inflow
regions of Fig.~\ref{fig:schematic}.  If the Larmor radius $\rho_{i2}$
of Population~2 ions significantly exceeds $\delta_{S2}$, it is not
clear how the basic dissipation region structure is maintained.
However, the scaling of basic reconnection properties in kinetic
simulations are consistent with the scaling theory based on fluid
conservation laws \cite{Malakit10}, so one does not expect a breakdown
of the conservation laws.  Instead, we conclude that an electric field
upstream of the dissipation region must appear to prevent Population~2
ions from crossing the stagnation point. This Larmor electric field arises
due to the premature leakage of a small percentage of Population~2 ions
into the dissipation region, which creates charge separation.

We now make this more quantitative.  From the above argument for its
existence, the spatial extent $\delta_{E,{\rm Larmor}}$ of the Larmor
electric field should scale with the Larmor radius $\rho_{i2}$ of
Population~2 ions,
\begin{equation}
  \delta_{E,{\rm Larmor}} \sim \rho_{i2}. \label{eqn:dELarmor}
\end{equation}
The magnitude of the electric field $E_{{\rm Larmor}}$ can be estimated by
noting that the premature leakage of Population~2 ions into the dissipation region
will occur until the potential barrier for an ion due to the Larmor electric field
is comparable to the average kinetic energy per charge:
\begin{equation}
  E_{{\rm Larmor}} \sim \frac{k_B T_{i2}}{e\rho_{i2}}, \label{eqn:Elarmor}
\end{equation}
where $T_{i2}$ is the temperature of Population~2 ions, $e$ is the ion
charge, and $k_{B}$ is Boltzmann's constant.  Note that the energy density
of this Larmor electric field is negligible compared to the thermal
energy density of the Population~2 plasma.
Finally, the Larmor electric field should only exist when
\begin{equation}
  \rho_{i2} > \delta_{S2}. \label{eqn:Econdition}
\end{equation}
In the remainder of this paper, we present evidence from simulations
for the Larmor electric field and discuss implications for
observations in the Earth's magnetosphere.

\section{\label{sec:sim}Simulations}

%{\bf Simulations}
We use the parallel particle-in-cell (PIC) code P3D \cite{Zeiler02} to
perform simulations in 2.5 dimensions of collisionless anti-parallel
asymmetric reconnection.
%The velocity of each particle is stepped
%forward using the relativistic form of Newton's second law with the
%Lorentz force being the only force acting on the particles.
%Collisions are not included.  Magnetic and electric fields are evolved
%using Faraday's law and the Amp\`ere-Maxwell law, respectively.
%Charge conservation is maintained by correcting the electric field
%with a multigrid solver.
In the simulations, magnetic field strengths and particle number
densities are normalized to arbitrary values $B_0$ and $n_0$,
respectively.  Lengths are normalized to the ion inertial length
$d_{i0} = c/\omega_{pi}$ at the reference density. Time is normalized
to the ion cyclotron time $\Omega^{-1}_{ci0}=(eB_0/m_ic)^{-1}$.
Speeds are normalized to the Alfv\'{e}n speed $c_{A0} = B_0/(4\pi
m_in_0)^{1/2}$.  Electric fields and temperatures are normalized to
$E_0=c_{A0}B_0/c$ and $T_0=m_ic_{A0}^2/k_{B}$, respectively.

Simulations are performed in a periodic domain of size $L_x \times L_y
= 204.8 \times 102.4$ with a grid scale $\Delta x = \Delta y$ of
0.05. The time step $\Delta t$ is 0.0025, 0.005, or 0.01.
The normalization density $n_0$ is represented by a number of
particles per grid cell, or $ppg$, which ranges from 50 to 200.  The
ion to electron mass ratio for all but Run~1* is $m_i/m_e = 25$ and the speed of light
$c$ is $15 c_{A0}$.

The initial conditions are a double asymmetric current sheet
%with tanh-shaped
%magnetic fields $B_{x}$ varying in the $y$-direction.  There is
%initially no out-of-plane (guide) magnetic field ($B_z=0$ at $t=0$).
%The variation of the temperature $T$ also has a tanh-shaped profile.
%The ratio of ion temperature to electron temperature $T_i/T_e$ is 2.
%The density $n$ varies across the current sheets such that the total pressure
%balance in the fluid sense is achieved (see Ref.~\cite{Malakit10} for
%more description of the equilibrium).
(see Ref.~\cite{Malakit10} for
more details).
%This is not a kinetic equilibrium, but it
%quickly relaxes.
A small magnetic perturbation is used to initiate
reconnection.  Each simulation is evolved until reconnection reaches a
steady state. The parameters (magnetic fields, densities, electron temperatures and ion
temperatures on either side of the dissipation region) for each
simulation are shown in Table~\ref{tab:asym2}.
%The simulations are separated into
%different categories given in the ``Run'' column.  An ``N'' or ``n''
%means the density is asymmetric, while a ``B'' or ``b'' means the
%magnetic field is asymmetric.  Both descriptors are included when both
%are asymmetric.  In these cases, the simulations are further
%distinguished by the predicted relative locations of the stagnation
%point and X-line using predictions from Ref.~\cite{Cassak07d}.  In
%particular, ``BN'' means the stagnation point is on the inflow side
%with the lower density and stronger magnetic field, ``bN'' means the
%stagnation point is on the inflow side with the lower density, ``Bn''
%means the stagnation point is on the inflow side with the stronger
%magnetic field, and ``BN0'' means the stagnation point is co-located
%with the X-line.

%\begin{table*}  %\hbox to \textwidth{\hfill
%\rotatebox{90}{%
%\begin{minipage}{\textheight}
\begin{table}
  \caption{\label{tab:asym2} Parameters for the present simulation
    study.  Subscripts ``1'' and ``2'' refer to the two upstream sides
    of the dissipation region.  The values give the magnetic field
    strengths $B$, number densities $n$, electron temperatures $T_{e}$
    and ion temperatures $T_{i}$. For each run, the width of the
    Larmor electric field $\delta_{E,Larmor}$ and
    its magnitude $E_{Larmor}$ are also provided.}
\begin{centering}
\begin{ruledtabular}
\begin{tabular}{c|cc|cc|cccc|cc}
%\multicolumn{12}{l}{S = symmetric, N = asymmetric density, B = asymmetric magnetic field}\tabularnewline
%%\multicolumn{12}{l}{N = asymmetric density, B = asymmetric magnetic field}\tabularnewline
%\multicolumn{12}{l}{BN, bN, Bn, BN0 = asymmetric density and magnetic field, where if using X-line as the reference,}\tabularnewline
%Run  & $B_{1}$  & $B_{2}$  & $n_{1}$  & $n_{2}$  & $T_{e1}$  & $T_{i1}$  & $T_{e2}$  & $T_{i2}$  & $\Delta x$  & $\Delta t$  & $ppg$ \tabularnewline
Run  & $B_{1}$  & $B_{2}$  & $n_{1}$  & $n_{2}$  & $T_{e1}$  & $T_{i1}$  & $T_{e2}$  & $T_{i2}$  & $\delta_{E,Larmor}$ & $E_{Larmor}$ \tabularnewline
\hline
%S-1  & 1.0  & 1.0  & 1.0  & 1.0  & 0.17  & 0.33  & 0.17  & 0.33  &  0.1  & 0.01  & 200 \tabularnewline
%%\hline
%S-2  & 1.0  & 1.0  & 1.0  & 1.0  & 0.67  & 1.33  & 0.67  & 1.33  &  0.1  & 0.01  & 200 \tabularnewline
%%\hline
%S-3  & 1.0  & 1.0  & 0.2  & 0.2  & 10.00 & 20.00 & 10.00 & 20.00 &  0.05 & 0.005 & 200 \tabularnewline
%%\hline
%S-4  & 1.0  & 1.0  & 0.5  & 0.5  & 10.00 & 20.00 & 10.00 & 20.00 &  0.05 & 0.005 & 100 \tabularnewline
%%\hline
%S-5  & 2.0  & 2.0  & 0.5  & 0.5  & 10.00 & 20.00 & 10.00 & 20.00 &  0.05 & 0.005 & 100 \tabularnewline
%%\hline
%BN-1  & 1.0 & 5.57 & 1.0  & 0.25 & 6.67  & 13.33 & 6.67  & 13.33 &  0.05 & 0.005 & 100 \tabularnewline
%%\hline
1  & 1.0  & 2.0  & 1.0  & 0.1  & 0.67  & 1.33  & 1.67  & 3.33 & 3.80 & 1.09 \tabularnewline
%\hline
1* & 1.0  & 2.0  & 1.0  & 0.1  & 0.67  & 1.33  & 1.67  & 3.33 & 3.50 & 0.86 \tabularnewline
%\hline
2  & 1.0  & 2.0  & 1.0  & 0.25  & 0.67 & 1.33  & 0.67  & 1.33 & 2.50 & 0.44 \tabularnewline
%\hline
3  & 1.0  & 1.0  & 1.0  & 0.1  & 0.67  & 1.33  & 6.67  & 13.33& 7.80 & 1.31 \tabularnewline
%\hline
4  & 2.0  & 2.0  & 1.0  & 0.1  & 0.67  & 1.33  & 6.67  & 13.33& 5.70 & 2.13 \tabularnewline
%\hline
5  & 1.0  & 1.0  & 1.0  & 0.5  & 0.67  & 1.33  & 1.33  & 2.67 & 4.15 & 0.19 \tabularnewline
%\hline
6  & 1.5  & 1.0  & 1.0  & 0.1  & 0.67  & 1.33  & 8.75  & 17.50& 8.10 & 1.25 \tabularnewline
%\hline
7  & 1.5  & 1.0  & 1.0  & 0.2  & 0.67  & 1.33  & 4.38  & 8.75 & 6.90 & 0.79 \tabularnewline
%\hline
8  & 1.5  & 1.0  & 1.0  & 0.44 & 0.67  & 1.33  & 1.99  & 3.98 & 5.35 & 0.37 \tabularnewline
%\hline
9  & 2.0  & 1.0  & 1.0  & 0.25 & 0.67  & 1.33  & 4.67  & 9.33 & 7.40 & 0.58 \tabularnewline
%\hline
10 & 3.0  & 1.0  & 1.0  & 0.11 & 0.67  & 1.33  & 18.03 & 36.03& 8.85 & 1.44 \tabularnewline
%\hline
11 & 2.0  & 1.0  & 1.0  & 0.44 & 0.67  & 1.33  & 2.65  & 5.30 & 4.20 & 0.42 \tabularnewline
%\hline
12 & 3.0  & 1.0  & 1.0  & 0.44 & 0.67  & 1.33  & 4.55  & 9.09 & 4.90 & 0.53 \tabularnewline
%\hline
13 & 1.5  & 1.0  & 1.0  & 1.0  & 0.46  & 0.92  & 0.67  & 1.33 & n/a  & 0.06 \tabularnewline
%\hline
14 & 2.0  & 1.0  & 1.0  & 1.0  & 0.17  & 0.33  & 0.67  & 1.33 & n/a  & 0.04 \tabularnewline
%\hline
15 & 2.0  & 1.0  & 0.1  & 0.1  & 0.67  & 1.33  & 5.67  & 11.33& n/a  & 0.09 \tabularnewline
%\hline
16 & 3.0  & 1.5  & 0.3  & 0.3  & 0.67  & 1.33  & 4.42  & 8.83 & n/a  & 0.07 \tabularnewline
%\hline
17 & 3.0  & 1.0  & 1.0  & 1.0  & 0.67  & 1.33  & 2.00  & 4.00 & n/a  & 0.09 \tabularnewline
%\hline
%\hline
%\hline
%\multicolumn{9}{l}{BN = stagnation point on inflow side with low $n$, high $B$}\tabularnewline
%\multicolumn{9}{l}{bN = stagnation point on inflow side with low $n$}\tabularnewline
%\multicolumn{9}{l}{Bn = stagnation point on inflow side with high $B$}\tabularnewline
%\multicolumn{9}{l}{BN0 = stagnation point co-located with the X-line}\tabularnewline
%\multicolumn{12}{l}{Note: The location of the stagnation is predicted following Cassak and Shay (PhPl, 2007)}\tabularnewline
\end{tabular}
%\par\end{centering}
\end{ruledtabular}
\end{centering}
\end{table}
%\end{minipage}}\hfill}
%\end{table*}

\section{\label{sec:results}Results and Discussion}

\begin{figure}
\centering{}\includegraphics[width=3.1in]{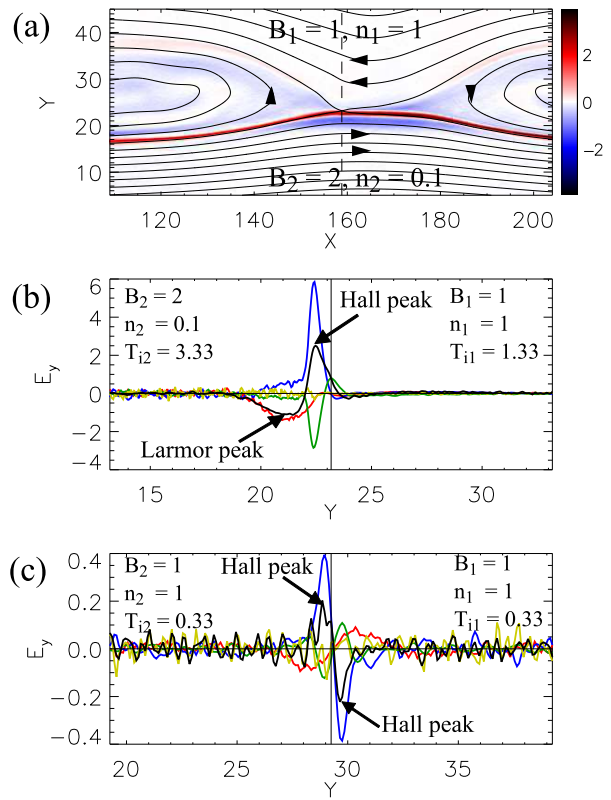}
\caption{\label{fig:ey} (a) Plot of the normal electric field $E_{y}$
  of Run~1, overplotted with magnetic field lines.  Red is
  positive, blue is negative, and white is zero.  (b) Cut of the
  normal electric field $E_{y}$ (black) and its contributing terms
  from generalized Ohm's law along the vertical dashed line in (a):
  the ion convection term $-(1/c)({\bf u_{i}}\times{\bf B})_{y}$
  (red), Hall term $(1/nec)({\bf J}\times{\bf B})_{y}$ (blue),
  pressure gradient term $-(1/ne)(\nabla\cdot{\bf P_{e}})_{y}$
  (green), and electron inertia term $-(m_{e}/e)(du_{e}/dt)_{y}$
  (yellow). (c) Similar cut to (b) but for data from a symmetric run.}
\end{figure}

We show results from Run~1 as a case study because it shows a
particularly clear example of the Larmor electric field and has
density and magnetic field variations typical of magnetopause reconnection.
The electric field $E_{y}$ in the inflow direction is shown in
Fig.~\ref{fig:ey}(a).  The blue band on the high density side and the
red band are the standard in-plane Hall electric field.  The blue band
on the side with the stronger magnetic field and lower density is the
Larmor electric field.  It is most prominent immediately upstream of
the dissipation region.

To show that this electric field is not caused by the Hall effect, we
present a plot of the contributions to $E_y$ from the generalized
Ohm's law in Fig.~\ref{fig:ey}(b) in a cut along the vertical dashed
line in Fig.~\ref{fig:ey}(a).  The vertical solid line marks the
X-line.  Immediately to the left of the X-line, there is an electric
field pointing toward the X-line (the positive peak).  This electric
field is largely contributed by the Hall term (in blue).  For comparison, 
Fig.~\ref{fig:ey}(c) shows a cut from a symmetric reconnection run, which has the 
standard bipolar structure of the Hall electric field on both sides 
of the X-line, and does not display a Larmor electric field.

Further toward smaller $y$ (to the left of the Hall field), there is
another electric field structure pointing in the opposite direction to
the Hall field.  However, the Hall term is very small in this region,
so it is not associated with the Hall term.  This electric field is
the Larmor electric field and points away from the X-line, in contrast
to the Hall electric field which points toward the X-line.  (For the
parameters of this simulation, the Hall electric field is skewed and
has a unipolar, rather than bipolar, electric field.) The Larmor
electric field occurs upstream of the dissipation region where the plasma
is roughly frozen-in, so it is roughly balanced by the ion convection
term in Ohm's law.

The appearance of this electric field is consistent with previous PIC
simulations of asymmetric reconnection \{see Fig~10(c) of
Ref.~\cite{Pritchett08} and Fig.~4(e) and 10(e) of Ref.~\cite{Tanaka08}\}.  It
may have been seen in observations \{see Fig.~9(h)) of
Ref.~\cite{Lindstedt09}\}.  However, its existence has not been
pointed out, and the physics of it has not yet been discussed.

\begin{figure}
\centering{}\includegraphics[width=3.1in]{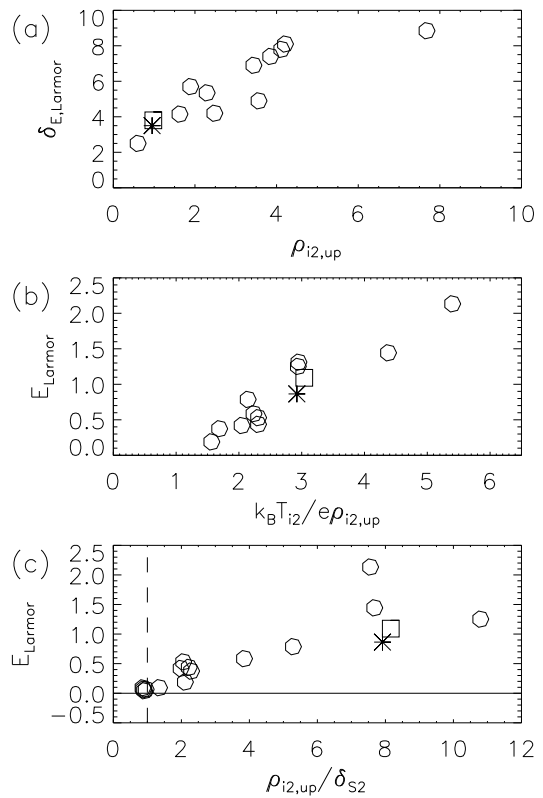}
\caption{\label{fig:key} (a) Thickness of the Larmor electric field
  $\delta_{E,{\rm Larmor}}$ vs.~ion Larmor radius $\rho_{i2}$ on the
  upstream side with the Larmor electric field.  (b) Magnitude of the
  Larmor field $E_{{\rm Larmor}}$ vs.~its predicted scaling
  $k_{B}T_{i2}/e\rho_{i2}$. (c) $E_{{\rm Larmor}}$ vs.~the ratio of
  $\rho_{i2}$ to the distance $\delta_{S2}$ between the stagnation
  point and the edge of the dissipation region. The square is for Run~1, and
  the asterisk is for Run~1*}
\end{figure}

We now consider the parametric dependence of properties of the Larmor
electric field.  For all simulations in Table~\ref{tab:asym2} for
which the Larmor electric field exists, we measure its spatial extent
$\delta_{E,{\rm Larmor}}$ in the inflow direction as the distance
between the two locations bounding the outward-directed electric
field, {\it i.e.,} where $E_{y}$ changes sign [at around $y=18.5$ and
22 in the example in Fig.~\ref{fig:ey}(b)].  The results are shown in
Fig.~\ref{fig:key}(a), where $\delta_{E,{\rm Larmor}}$ is plotted
versus the Larmor radius of the upstream ions $\rho_{i2}$, which is
evaluated using the ion temperature and magnetic field 10 $d_{i0}$
upstream of the X-line. A clear correlation between the two length scales is
apparent, in agreement with Eq.~(\ref{eqn:dELarmor}).

The magnitude of $E_{{\rm Larmor}}$ is measured where the electric field peaks
in the direction away from the X-line.  For simulations in which the
Larmor electric field exists, the results are plotted in
Fig.~\ref{fig:key}(b), with the predicted Larmor field strength $k_{B}
T_{i2} / e \rho_{i2}$ from Eq.~(\ref{eqn:Elarmor}) on the horizontal
axis.  Scaling according to Eq.~(\ref{eqn:Elarmor}) is observed.

This result suggests that the energy gained/lost by an ion that crosses
the electric field structure is only in the order of the thermal energy of an ion from Population~2 $k_{B}T_{i2}$.
Hence, this electric field does not significantly participate in particle acceleration.

Finally, we note that there are five simulations for which the Larmor
electric field does not arise.  From Eq.~(\ref{eqn:Econdition}), we
expect the Larmor electric field to exist when $\rho_{i2}$ exceeds the
distance $\delta_{S2}$ from the stagnation point to the edge of the
dissipation region on the side of the dissipation region with the
Larmor electric field.

To test this hypothesis, we must find $\delta_{S2}$.  This is
difficult to measure directly from the simulations due to the noise
inherent in PIC simulations, so we appeal to asymmetric reconnection
theory.  Equation~(\ref{eqn:ds2ds1}) gives the relative length scales,
so we need the total thickness of the dissipation region.  For
collisionless asymmetric reconnection, the half thickness of the
dissipation region $\delta$ was predicted to be \cite{Cassak09}
\begin{equation}
  \delta \sim \frac{B_{1}+B_{2}}{2\sqrt{B_{1}B_{2}}} \left[\left(
      \frac{m_{i}^{2}c^{2}}{4\pi e^{2}}\right)
    \left(\frac{B_{1}+B_{2}}{m_{i}(n_{1}B_{2}+n_{2}B_{1})}\right)\right]^{1/2}.
  \label{eqn:delta}
\end{equation}
Using $2 \delta = \delta_{S1} + \delta_{S2}$, one finds the predicted
absolute size of $\delta_{S2}$:
\begin{equation}
  \delta_{S2} \sim \left[\frac{n_2 B_1}{n_1 B_2+n_2 B_1}\right]2\delta.
  \label{eqn:ds2}
\end{equation}
For comparison with the simulations the values are calculated from the asymptotic values in
Table~\ref{tab:asym2}.

The results are plotted in Fig.~\ref{fig:key}(c), where the magnitude
of the Larmor electric field $E_{{\rm Larmor}}$ is plotted versus the
ratio $\rho_{i2}/\delta_{S2}$.  One sees that $E_{{\rm Larmor}}$ is
non-zero when $\rho_{i2}/\delta_{S2} \gtrsim 1$ and is near 0
otherwise, consistent with Eq.~(\ref{eqn:Econdition}). Note that the small
deviations of $E_{{\rm Larmor}}$ from zero for Runs~13-17 are merely due to noise,
and $\delta_{E,{\rm Larmor}}$ cannot sensibly be determined.

To test the effect of the electron mass, we perform a simulation like Run~1 with an ion to electron mass ratio of 100 instead of 25, which we call Run~1*. Comparison between Run~1 and Run~1* suggests that the results discussed above
are insensitive to the ion to electron mass ratio in the simulations. We therefore expect the results
to still hold for the actual mass ratio even though it is significantly larger.

%\section{\label{sec:summary}Summary}

%{\bf Summary}

\section{\label{sec:applications}Applications}

%We have argued that there can exist an in-plane electric
%field, not related to Hall physics, that points away from the X-line
%during collisionless asymmetric reconnection.  This electric field is
%associated with finite Larmor radius effects of the inflowing ions, so
%we dub it the ``Larmor electric field.''  Using theoretical arguments
%and fully kinetic PIC simulations of collisionless asymmetric
%reconnection with various inflow conditions, we have shown that the
%Larmor electric field occurs when the Larmor radius of the upstream
%ions is larger than the distance from the stagnation point to the edge
%of the dissipation region, has a spatial extent in the inflow
%direction comparable to the upstream ion Larmor radius, and has a
%magnitude scaling as $E_{{\rm Larmor}}\sim k_B T_{i2}/e \rho_{i2}$.

For reconnection at the dayside magnetopause, we argue that the Larmor
electric field is expected to be present.  Typical magnetic field
strengths, densities, and ion temperatures on the magnetosheath side
are approximately 20 nT, 25 cm$^{-3}$, and $2\times10^{6}$ K, and are
55 nT, 0.5 cm$^{-3}$, and $20\times10^{6}$ K on the magnetospheric
side \cite{Phan96}.  The distances from the stagnation point to the
edges of the dissipation region are calculated using
Eqs.~(\ref{eqn:delta}) and (\ref{eqn:ds2}) to be about 1 km on the
magnetospheric side and 119 km on the magnetosheath side.  The ion
Larmor radii on the magnetospheric and magnetosheath sides are 77 km
and 67 km. Since the ion Larmor radius is expected to be larger than
the distance from the stagnation point to the edge of the dissipation
region on the magnetospheric side, the Larmor electric field is
expected to exist on the magnetospheric side pointing away from the
X-line, {\it i.e.,} toward the Earth.  The magnitude of the electric
field, from Eq.~(\ref{eqn:Elarmor}), is predicted to be on the order
of 20 ${\rm mV/m}$.  This strength of electric field is easily
measurable by spacecraft.

Since the Larmor electric field is localized upstream of the
dissipation region, it can be a useful signature to help satellites,
such as the MMS mission, identify the dissipation region of
reconnection sites before the satellite moves deeper into the
dissipation region.

\begin{acknowledgments}
  This research was supported by the postdoctoral research sponsorship
  of Mahidol University (KM), NSF grants ATM-0645271 - Career Award (MAS) and AGS-0953463 (PAC),
  NASA grants NNX08A083G - MMS IDS, NNX11AD69G, and NNX13AD72G (MAS), NNX10AN08A (PAC),
  and the Thailand Research Fund (DR).
  Simulations were performed at the National
  Center for Atmospheric Research Computational and Information System
  Laboratory (NCAR-CISL).  The authors thank Yu.~Khotyaintsev for
  pointing out the observations and M.~M.~Kuznetsova for her helpful discussion.
\end{acknowledgments}

\bibliography{apssamp}% Produces the bibliography via BibTeX.

\end{document}